\documentclass[graybox]{svmult}

\usepackage{mathptmx}       
\usepackage{helvet}         
\usepackage{courier}        
\usepackage{type1cm}        
\usepackage{makeidx}         
\usepackage{graphicx}        
\usepackage{multicol}        
\usepackage[bottom]{footmisc}

\usepackage{bm}
\usepackage{array}
\usepackage{mhchem}
\usepackage{amsfonts}

\makeindex             

\newcommand{\icm}{cm$^{-1}$}

\newcommand{\kcalA}{kcal~mol$^{-1}$~\AA$^{-1}$}

\begin{document}

\title*{Molecular Dynamics with Neural-Network Potentials}
\author{Michael Gastegger and Philipp Marquetand}
\institute{Michael Gastegger\at Technical University of Berlin, Machine Learning Group, Marchstr. 23, 10587 Berlin, Germany. \email{michael.gastegger@tu-berlin.de}
\and Philipp Marquetand \at University of Vienna, Faculty of Chemistry, Institute of Theoretical Chemistry, W\"ahringer Str. 17, 1090 Vienna, Austria. \email{philipp.marquetand@univie.ac.at}}
%
%
\maketitle

\abstract*{Molecular dynamics simulations are an important tool for describing the evolution of a chemical system with time. However, these simulations are inherently held back either by the prohibitive cost of accurate electronic structure theory computations or the limited accuracy of classical empirical force fields. Machine learning techniques can help to overcome these limitations by providing access to potential energies, forces and other molecular properties modeled directly after an electronic structure reference at only a fraction of the original computational cost. The present text discusses several practical aspects of conducting machine learning driven molecular dynamics simulations.
First, we study the efficient selection of reference data points on the basis of an active learning inspired adaptive sampling scheme.
This is followed by the analysis of a machine-learning based model for simulating molecular dipole moments in the framework of predicting infrared spectra via molecular dynamics simulations.
Finally, we show that machine learning models can offer valuable aid in understanding chemical systems beyond a simple prediction of quantities.
}

\abstract{Molecular dynamics simulations are an important tool for describing the evolution of a chemical system with time. However, these simulations are inherently held back either by the prohibitive cost of accurate electronic structure theory computations or the limited accuracy of classical empirical force fields. Machine learning techniques can help to overcome these limitations by providing access to potential energies, forces and other molecular properties modeled directly after an accurate electronic structure reference at only a fraction of the original computational cost. The present text discusses several practical aspects of conducting machine learning driven molecular dynamics simulations.
First, we study the efficient selection of reference data points on the basis of an active learning inspired adaptive sampling scheme.
This is followed by the analysis of a machine-learning based model for simulating molecular dipole moments in the framework of predicting infrared spectra via molecular dynamics simulations.
Finally, we show that machine learning models can offer valuable aid in understanding chemical systems beyond a simple prediction of quantities.
}

\section{Introduction}
\label{sec:1}
Chemistry is -- in large part -- concerned with the changes that matter undergoes. As such, chemistry is inherently time-dependent and if we want to model such chemical processes, then a time-dependent approach is most intuitive. The corresponding techniques can be summarized as dynamics simulations. In particular molecular dynamics (MD) simulations -- defined usually as treating the nuclear motion with Newton's classical mechanics -- are commonly used to "mimic what atoms do in real life" \cite{Marx2009}. Such simulations have become indispensable not only in chemistry but also in adjacent fields like biology and material science \cite{Allen1987, Frenkel2001}. 

An important ingredient of MD simulations are molecular forces, which determine how the nuclei move. For the sake of accuracy, it is desirable to obtain these forces from quantum-mechanical electronic structure calculations. However, such \emph{ab initio} calculations are expensive from a computational perspective and hence only feasible for relatively small systems or short timescales. For larger systems (e.g. proteins) molecular forces are instead modeled by classical force fields, which are composed of analytic functions based on physical findings. As a consequence of these approximations, classical force fields are extremely efficient to compute but fail to reach the accuracy of electronic structure methods.

With the aim to obtain both a high accuracy and a fast evaluation, machine learning (ML) is employed to predict forces and an increasing number of research efforts are devoted to this idea. Since the forces are commonly evaluated as the derivative of the potential energy with respect to the coordinates, the learning of energies and forces are tightly connected and we often find the term ''machine learning potential'' in the literature. 

Possibly the first work to use ML of potentials combined with dynamics simulations appeared in 1995 by Blank \emph{et al.} \cite{Blank1995JCP}, where diffusion of CO on a Ni surface and H$_2$ on Si was modelled. A comprehensive overview over the earlier work in this field, where mostly neural networks were employed, can be found in the reviews \cite{Latino2010IJQC, Behler2011PCCP}. Later, also other ML methods like Gaussian approximation potentials \cite{Bartok2010PRL} were utilized, diversifying the research landscape, which is reflected in various, more topical reviews, see e.g. \cite{Jiang2016IRPC, Ramakrishnan2017, Botu2017JPCC, Behler2017ACIE}.
Today, the field has become so active, that a review would be outdated by tomorrow. Here instead, we give an example of what can be achieved when combining ML and MD.

In this chapter, we describe simulations of one of the most fundamental experiments to detect moving atoms, namely infrared spectra. The simulations utilize MD based on potentials generated with high-dimensional neural networks, a special ML architecture. We show in particular, how the training data can efficiently be gathered by an adaptive sampling scheme. Several practical aspects, tricks and pitfalls are presented. Special emphasis is put also on the prediction of dipole moments and atomic charges, which are necessary ingredients besides potential energies and forces for the calculation of infrared spectra from MD.


\section{Methods}
\label{sec:2}

\subsection{High-dimensional neural network potentials}
\label{subsec:2.1}

\begin{figure}
	\centering
	\includegraphics[width=0.7\textwidth]{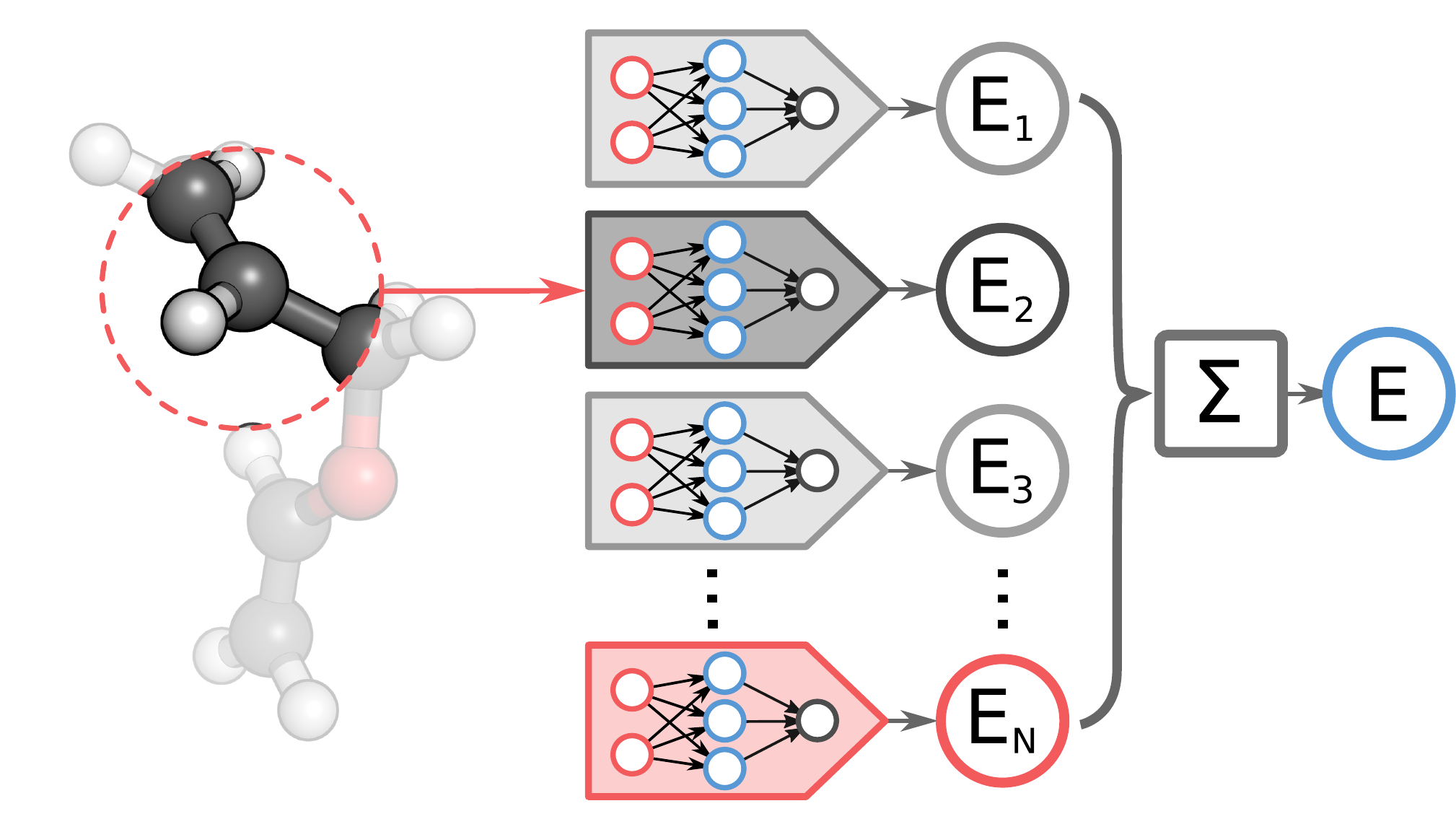}
	\caption{In a high-dimensional neural network potential, the local chemical environment of each atom is first encoded in a structural descriptor. Based on these descriptors, neural networks predict atomistic energy contributions, where different networks are used for the different chemical elements. Finally, the atomic energies are summed in order to recover the total energy of the system.}
	\label{fig:hdnn}
\end{figure}

High-dimensional neural network potentials (NNPs) are a type of atomistic ML potentials \cite{Behler2007PRL}. Atomistic potentials model the properties of a system based on the contributions of individual atoms due to their local chemical environment (Figure~\ref{fig:hdnn}).
In high-dimensional NNPs, atomic environments are represented via so-called atom-centered symmetry functions (ACSFs) \cite{Behler2011JCP}. Typically, ACSFs are radial and angular distribution functions, which account for rotational and translation invariances of the system. A radial symmetry function, for example, is a superposition of Gaussian densities:
\begin{equation}
G^\mathrm{rad}_i = \sum_{j\neq i}^N e^{-\eta(r_{ij}-r_0)^2} f_\mathrm{cut}(r_{ij}).
\end{equation}
The sum includes all atoms $j$ in vicinity of the central atom $i$. $r_{ij}$ is the distance between $i$ and $j$. $\eta$ and $r_0$ are parameters which modulate the width and center of the Gaussian. A cutoff function $f_\mathrm{cut}$ ensures, that only the local chemical environment contributes to the ACSF. For a more in-depth discussion on ACSFs and their features, we refer to References~\cite{Behler2011JCP,Behler2015IJQC,Behler2017ACIE,Gastegger2018JCP}.

Based on the ACSF representation of each atom, an atomistic neural network then predicts the contribution of this atom to the global molecular property. Finally, these contributions are recombined via an atomistic aggregation layer in order to recover the target property. For NNPs which model the potential energy $E$ of a system, this layer is usually chosen as a sum over the individual atomic energies $E_i$:
\begin{equation}
\tilde{E} = \sum^N_i \tilde{E}_i, \label{eq:atpot}
\end{equation}
where $N$ is the number of atoms in the molecule. 
However, different aggregation layers can be formulated in order to model various properties, as will be discussed in the next section.

Due to the form of high-dimensional NNPs, expressions for analytic Cartesian derivatives of the model are readily available. Hence, NNPs provide access to energy conserving forces
\begin{equation}
\mathbf{\tilde{F}}_{\alpha} = -\sum^N_i \sum^{D_i}_d \frac{\partial \tilde{E}_i}{\partial G_d} \frac{\partial G_d}{\partial \mathbf{R}_{\alpha}}, \label{eq:forces}
\end{equation}
where $\mathbf{\tilde{F}}_{\alpha}$ and $\mathbf{R}_{\alpha}$ are the forces acting on atom $\alpha$ and its position, while $D_i$ is the number of ACSFs centered on atom $i$. NNP forces can also be included into the training process by minimizing a loss function of the form
\begin{equation}
\mathcal{L} = \frac{1}{M} \sum^M_m \left( \tilde{E}_m - E_m \right)^2  + \frac{\vartheta}{M} \sum^{M}_m \frac{1}{3N_m} \sum^{N_m}_i \left\| \widetilde{\mathbf{F}}_{mi} - \mathbf{F}_{mi} \right\|^2. \label{eq:lossforc}
\end{equation}
Here, $M$ is the number of molecules present in the data set. $\vartheta$ controls the trade-off between fitting energies and forces, while $\mathbf{F}_{mi}$ is the vector of Cartesian forces acting on atom $i$ of molecule $m$.
By using forces during training, $3N$ additional pieces of information are available for each molecule beside the potential energy. As a consequence, the overall number of reference computations required to construct an accurate NNP can be reduced significantly \cite{Pukrittayakamee2009JCP,Gastegger2015JCTC}.


\subsection{Dipole model}
\label{subsec:2.2}
 
In addition to energies and forces, atomistic aggregation layers can also be formulated to model various other molecular properties.
One example is the dipole moment model introduced in Reference~\cite{Gastegger2017CS}. Here, the dipole moment $\bm{\mu}$ of a molecule is expressed as a system of atomic point charges, according to the relation
\begin{equation}
\tilde{\bm{\mu}} = \sum^N_i \tilde{q}_i \mathbf{r}_i. \label{eq:atmu}
\end{equation}
$\tilde{q}_i$ is the charge located at atom $i$ and $\mathbf{r}_i$ is the position vector of the atom relative to the molecules center of mass. 
The charges $\tilde{q}$ are modeled via atomistic networks and depend on the local chemical environment. However, these point charges are never learned directly, but instead represent latent variables inferred by the NNP dipole model during training, where the following loss function is minimized: 
\begin{equation}
\mathcal{L} = \frac{1}{M} \sum^M_m \left( \tilde{Q}_m - Q_m \right)^2 + \frac{1}{3M} \sum^M_m \left\| \tilde{\bm{\mu}}_{m} - \bm{\mu}_{m} \right\|^2 + \ldots \label{eq:qnn}
\end{equation}
$\bm{\tilde{\mu}}$ is the expression for the dipole moment given in Equation~\ref{eq:atmu} and $\bm{\mu}$ is the electronic structure reference. The first term in the loss function introduces the additional constraint, that the sum of latent charges $\tilde{Q} = \sum_i \tilde{q}_i$ should reproduce the total charge $Q$ of the molecule.
Formulated in this way, the machine learning model depends only on quantum mechanical observables in the form of a molecule's electrostatic moments (total charge and dipole moment).

Expression~\ref{eq:qnn} can easily be extended to include higher moments such as the quadrupole moment $\bm{\Theta}$, as was suggested in Reference~\cite{Gastegger2017CS}. In the context of the above model, $\bm{\Theta}$ takes the form
\begin{equation}
\bm{\tilde{\Theta}} = \sum_i \tilde{q}_i \left( 3 \mathbf{r}_i \otimes \mathbf{r}_i - \mathbf{1} \left\| \mathbf{r}_i  \right\| \right),
\end{equation}
where, $\mathbf{r}_i \otimes \mathbf{r}_i$ is the outer product of the Cartesian position vectors of atom $i$. 
However, it was found that the introduction of quadrupole moments offers no additional advantage, at least when modeling dipoles.
Moreover, using an atomistic model for $\bm{\Theta}$ can be problematic for small molecules such as water, since the atom centered point charges are not able to resolve features of the charge distribution arising from e.g. the lone pair electrons of the oxygen.

It should be emphasized at this point, that the atomistic aggregation layers presented here are not restricted to a single type of machine learning architecture. They can be coupled with any model in a modular fashion, as long as it provides access to atomic contributions. This was for example done recently with the SchNet architecture in order to model dipole moment magnitudes \cite{Schuett2018arXiv}.

\subsection{Adaptive Sampling Scheme}
\label{subsec:2.3}

Before a NNP can be used for simulations, its free parameters need to be determined by training on a suitable set of reference data.
Typically, a set of reference molecules is chosen in a two-step process.
First, the PES is sampled to obtain a representative set of molecular configurations.
Afterwards, the quantum chemical properties of these structures (e.g. energies, forces, \ldots) are computed with an appropriate electronic structure method.
The sampling can be performed in different ways, with molecular dynamics and normal mode sampling being only a few examples \cite{Behler2015IJQC,Smith2017CS}.
However, a feature shared by most sampling methods is, that they either use approximate methods such as molecular force fields to guide the sampling or they perform all simulations at the final level of theory.
Both approaches have drawbacks. In the first case, the PES regions explored with the lower level of theory need not correspond to regions relevant for the high level model (e.g. different molecular geometries).
In the second case, the unfortunate scaling of accurate electronic structure methods limits either the regions of the PES that can be covered or the overall accuracy of the reference computations.

A solution to these issues is to use an adaptive sampling scheme, where the ML model itself selects new data points to improve its accuracy \cite{Gastegger2017CS}. This approach is inspired by active learning techniques and proceeds as follows (Figure~\ref{fig:samp}):
\begin{figure}
	\centering
	\includegraphics[width=0.7\textwidth]{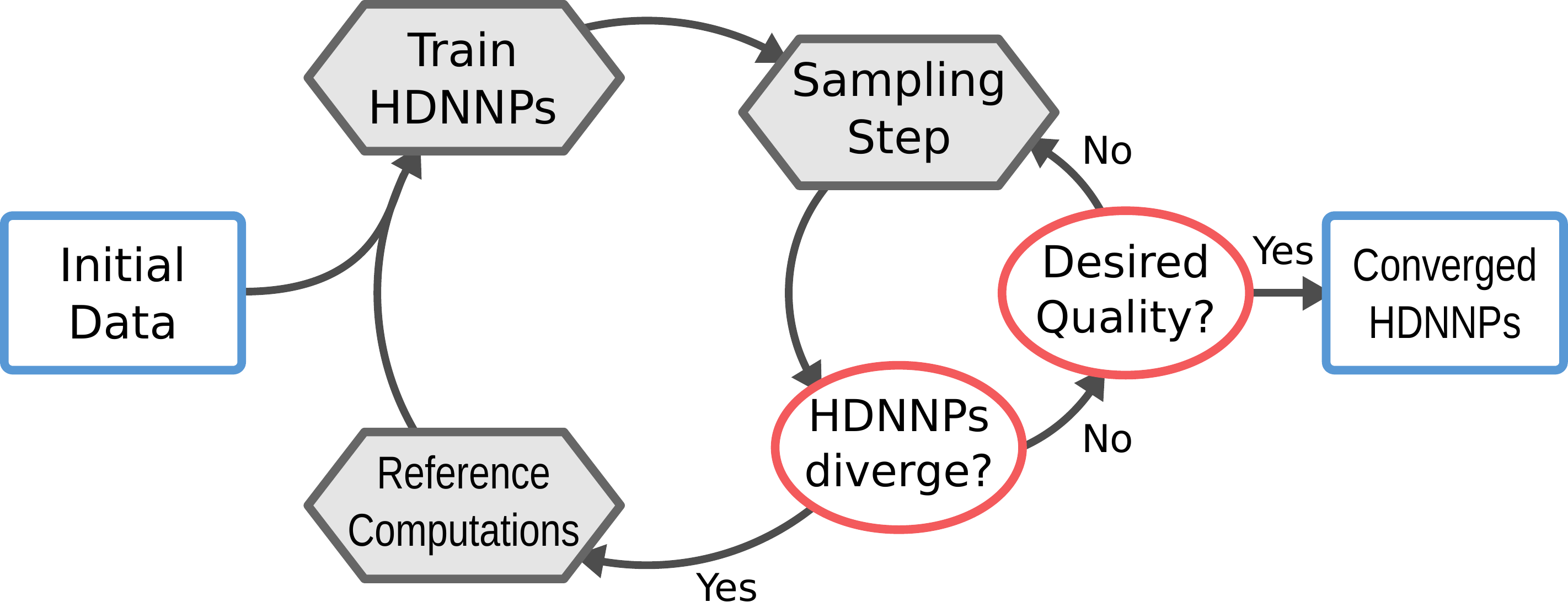}
	\caption{The adaptive sampling scheme starts by training a preliminary ensemble of NNPs on a small set of reference computations. This ensemble is then used to sample new configurations via e.g. molecular dynamics. For every sampled configuration, an uncertainty measure is computed. If this measure exceeds a predefined threshold, the simulation is stopped and a reference calculation is performed. The new data point is then added to the reference data and the next generation of NNPs is trained. This procedure is repeated in an iterative manner until a convergence threshold is reached.}
	\label{fig:samp}
\end{figure}
First, a crude NNP model is used to explore an initial region of the PES. During simulation, an uncertainty measure for the NNP predictions is monitored. If this measure exceeds a threshold, the sampling is stopped and electronic structure computations are performed for the corresponding configuration. The resulting data is then added to the reference set and used to update the NNP. Sampling is continued with the improved model. This procedure is carried out in an iterative fashion until the desired level of accuracy is reached.

One advantage of this scheme is that the NNP model used to guide the sampling closely resembles the electronic structure reference method for large stretches of coordinate space. Hence, similar regions of the PES will be explored (e.g. bond lengths) as if the simulations were carried out with the reference method exclusively.

In addition, by using the model uncertainty to determine when additional reference computations should be performed, only a small number of expensive calculations are necessary.
Due to the simplicity of this scheme, it can easily be combined with different sampling methods, such as molecular dynamics, metadynamics or Monte-Carlo based approaches \cite{Herr2018JCP,Smith2018JCP}.

Perhaps the most important ingredient for the above scheme is an appropriate uncertainty measure. Here, it is possible to make use of a trait of NNs in general and NNPs in particular.
Two NNPs trained on the same reference data will agree closely for regions of the PES described well by both models. However, in regions sampled insufficiently the predictions of both models will diverge quickly.
Using the disagreement between different models to guide data selection is a popular approach in active learning called query by committee \cite{Seung1992}.
Based on the above behavior, one can formulate the following uncertainty measure for NNPs:
\begin{equation}
\sigma_E = \sqrt{ \frac{1}{\mathfrak{N}-1} \sum^\mathfrak{N}_\mathfrak{n} \left( \tilde{E}_\mathfrak{n} - \overline{E} \right)^2}.\label{eq:unc}
\end{equation}
$\tilde{E}_\mathfrak{n}$ is the energy predicted by one of $\mathfrak{N}$ different NNPs and $\overline{E}$ is the average of all predictions. Hence, $\sigma_E$ is the standard deviation of the different model predictions.
Using an uncertainty measure of this form also has the following advantage:
Since different NNPs are used to compute $\sigma_E$, they can be combined into an ensemble, where the prediction averages (e.g. $\overline{E}$) are used to guide PES exploration. The consequence is an improvement in the overall predictive accuracy of the ML approach at virtually no extra cost, due to error cancellation in the individual models.


\section{Generation of Reference Data Sets}
\label{sec:3}

The following section discusses different practical aspects of the adaptive sampling scheme introduced above.
After an investigation on the accuracy advantage offered by NNP ensembles, we study how frequently new reference computations are requested during a sampling run. Afterwards, the utility of different predicted properties as uncertainty measures for NNPs is analyzed.
Finally, we introduce an extension to the standard sampling scheme, which improves overall sampling efficiency.

\subsection{Accuracy of NNP Ensembles}
\label{subsec:3.2}

\begin{figure}
	\centering
	\includegraphics[width=\textwidth]{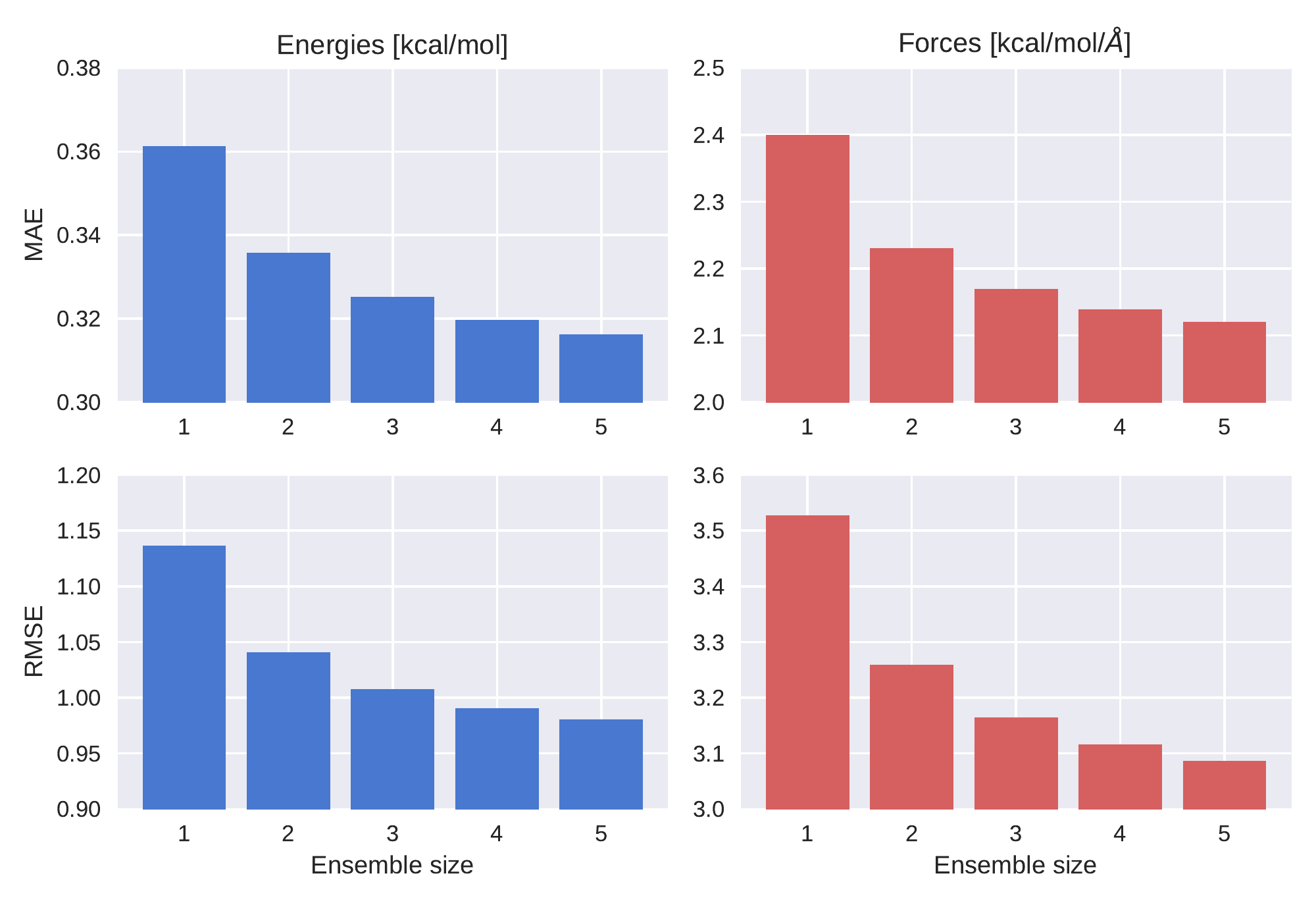}
	\caption{Accuracy of ensemble predictions for molecular energies (blue) and forces (red) depending on the number of members. The computed error measures, MAE and RMSE, appear to decrease according to an $\frac{1}{\sqrt{\mathfrak{N}}}$ relation, where $\mathfrak{N}$ is the number of models in the ensemble. In all cases, the gain in accuracy is most pronounced when going from a single network to an ensemble of two.}
	\label{fig:ens}
\end{figure}

In order to investigate the accuracy offered by ensembles of NNPs, we compare the predictions of ensembles containing up to five members to their respective electronic structure reference.
This analysis is based on the protonated alanine tripetide data set obtained in Reference~\cite{Gastegger2017CS}, which also serves as a basis for several other studies in this text.
The data set contains 718 different peptide configurations at the BLYP/def2-DZVP level of theory sampled with the scheme described above.
Figure~\ref{fig:ens} shows the energy and force mean absolute errors (MAEs) and root mean squared errors (RMSEs) for the different ensembles.
Even the combination of only two different models already leads to a marked decrease in the prediction error.
Since ensembles thrive on a cancellation of random error fluctuations, this gain in accuracy is particularly pronounced for the RMSEs.
An interesting observation is that the forces profit to a greater extent than the energies, with a reduction in the error by approximately 0.3~\kcalA.
This effect is expected to be of importance in the early stages of an adaptive sampling run, as the improved reliability of the model increases the likelihood that physically relevant configurations are sampled.

\subsection{Frequency of Reference Computations}
\label{subsec:3.1}

\begin{figure}
	\centering
	\includegraphics[width=0.8\textwidth]{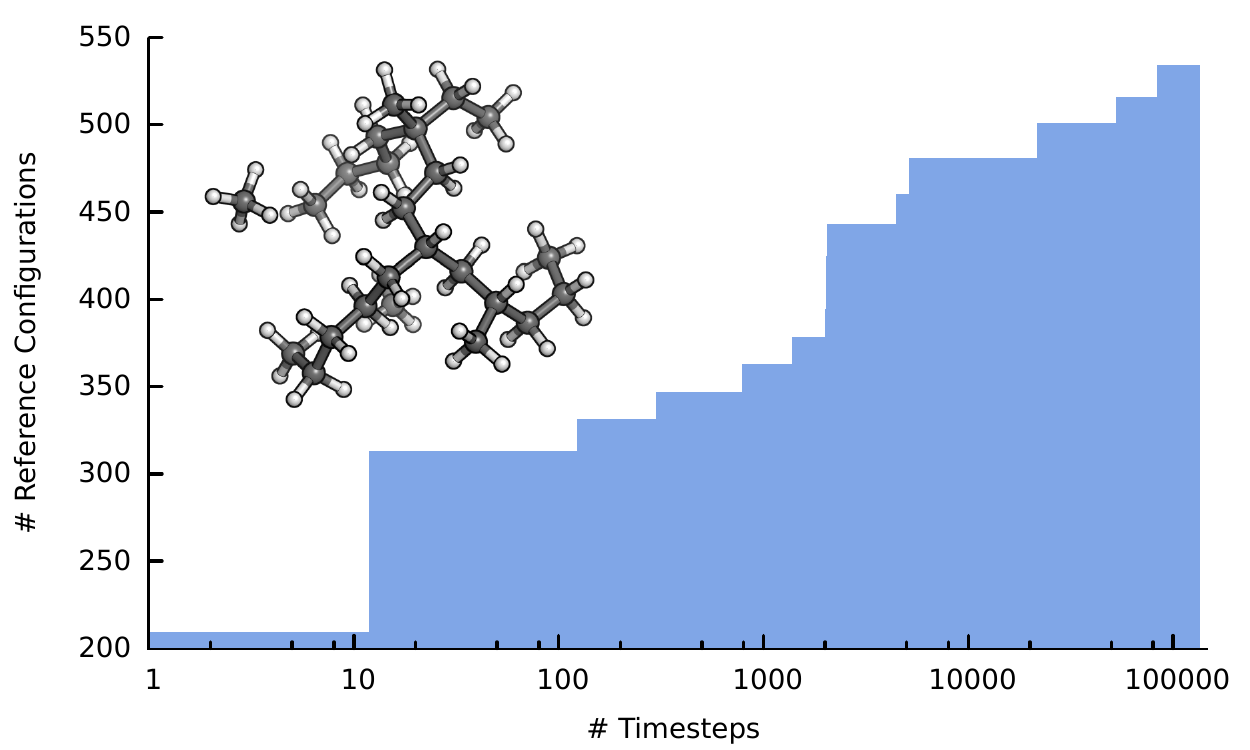}
	\caption{Number of configurations accumulated during an adaptive sampling run for the \ce{C69H140} n-alkane plotted against the number of molecular dynamics steps. New samples are added frequently during the early stages of the sampling, while almost no configurations are collected during the later stages.}
	\label{fig:samp_alk}
\end{figure}

An important aspect of the adaptive sampling scheme is how frequently new electronic structure computations need to be performed.
Figure~\ref{fig:samp_alk} depicts the number of configurations added to the reference data set versus the total number of sampling steps.
The studied molecule is an n-alkane (\ce{C69H140}, see Figure inset) and the sampling statistics were taken from the supporting information of Reference~\cite{Gastegger2017CS}.
As can be seen in the figure, there is a marked decrease in the number of electronic structure queries as the sampling progresses.
Initially, new samples are added frequently, as the model explores the configuration space. More than half of the new samples are added within the first 2000 exploration steps. After this phase of determining a reliable first approximation of the electronic structure method, the sampling process is dominated by fine-tuning the NNP ensemble. Now only samples corresponding to insufficiently described regions of the PES are collected, reducing the requirement for expensive reference computations significantly.
The efficiency of this simple approach is remarkable insofar, as only 534 configurations are needed to obtain an accurate model of the n-alkane sporting 621 degrees of freedom.

\subsection{Choice of Uncertainty Measures}
\label{subsec:3.3}

An important feature of atomistic NNPs is their ability to operate as fragmentation approaches, where they predict the energies of large molecules after being trained on only small fragments \cite{Gastegger2016JCP}.
Hence, expensive reference computations never have to be performed for the whole system, but only for parts of it. 
This feature can be combined with the adaptive sampling scheme, as was for example done in Reference~\cite{Gastegger2017CS}. In this setup, the uncertainty is not measured for the whole molecule, but instead for atom-centered fragments.
Reference computations are only performed for those fragments where the uncertainty exceeds a predefined threshold, thus reducing the computational cost of constructing an accurate NNP even further. However, the deviation of ensemble energies (Equation~\ref{eq:unc}) can now no longer serve as the uncertainty measure. 

\begin{figure}
	\centering
	\includegraphics[width=0.9\textwidth]{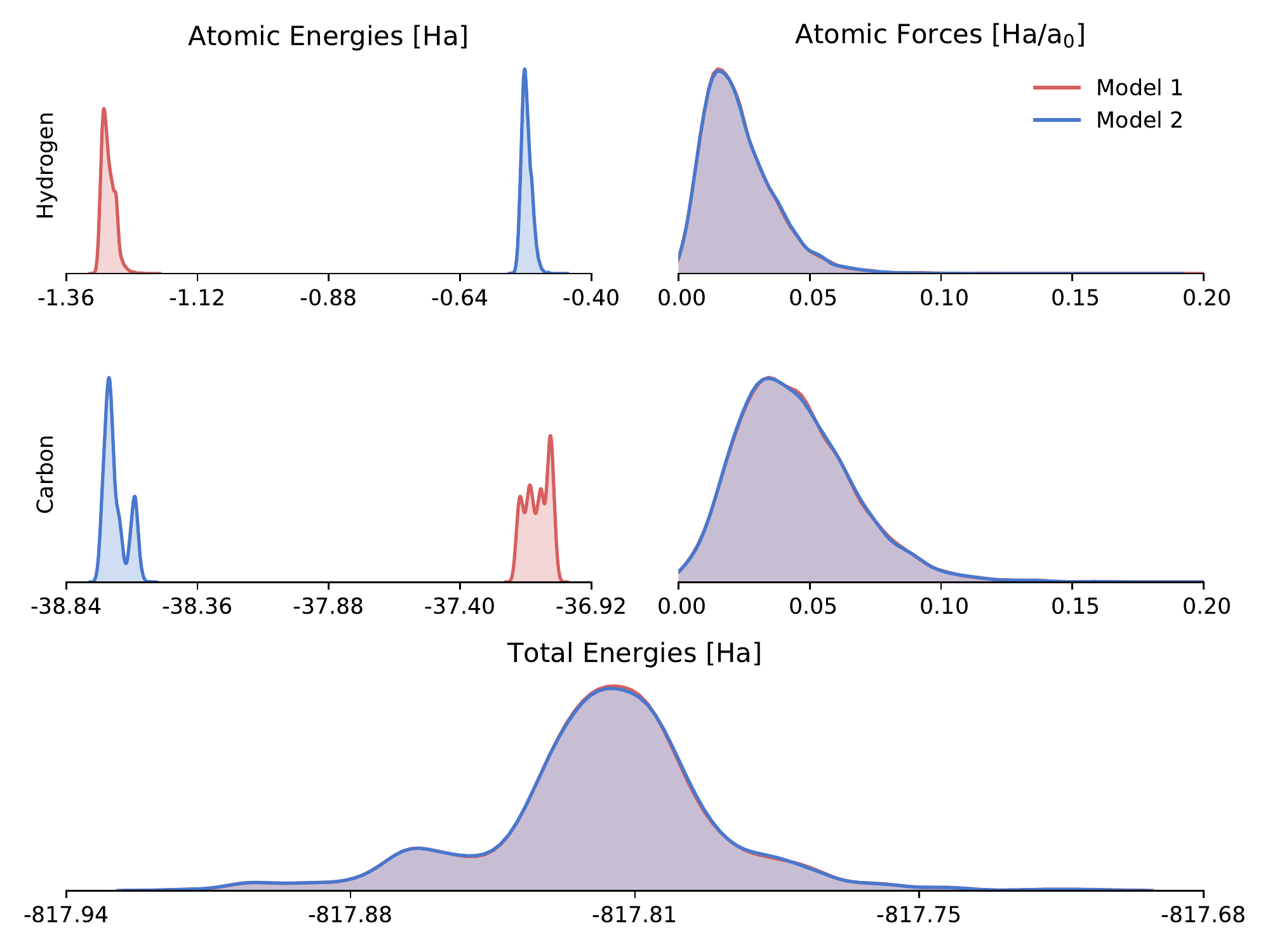}
	\caption{Distribution of atomic energies, forces and total energies as predicted for the alanine tripeptide by two NNP models (shown in red and blue). Although the NNP predictions agree well in the case of the total energies and atomic forces, the energy contributions of individual atoms vary dramatically between the models.}
	\label{fig:prop_dist}
\end{figure}

Although substituting the total energies in Equation~\ref{eq:unc} by their atom-wise counterparts $\tilde{E}_i$ would in theory be a straightforward choice for an atomistic uncertainty estimate, it is not feasible in practice.
Due to the way NNPs are constructed (see Equation~\ref{eq:atpot}), the partitioning of the total energy into latent atomic contributions is not unique.
Hence, even if two NNPs yield almost identical predictions for the molecular energies, the distributions of atom-wise contributions can still differ significantly, as is shown for the alanine tripetide in Figure~\ref{fig:prop_dist}.
If e.g. the atomic energies of carbon atoms are used to compute the uncertainty, large deviations will be encountered for all regions of the PES, no matter how well the global predictions agree. As a consequence, reference computations will be performed for a large fraction of encountered configurations, thus effectively negating the advantage offered by the adaptive sampling scheme.

The better alternative is to reformulate the above measure to instead use the forces acting on each atom:
\begin{equation}
\sigma_F^{(\alpha)} = \sqrt{ \frac{1}{\mathfrak{N}-1} \sum^\mathfrak{N}_\mathfrak{n} \left\|  \tilde{\mathbf{F}}^{(\alpha)}_\mathfrak{n} - \overline{\mathbf{F}}^{(\alpha)}  \right\|^2},
\end{equation}
where $\tilde{\mathbf{F}}^{(\alpha)}_\mathfrak{n}$ is the force acting on atom $\alpha$ as predicted by model $\mathfrak{n}$ of the ensemble. $\overline{\mathbf{F}}^{(\alpha)}$ is the average over all predictions. The measure $\sigma_F^{(\alpha)}$ has several advantages. Since it depends on the molecular forces it is purely atomistic. Moreover, due to how the forces are computed in NNPs (Equation~\ref{eq:forces}), they are insensitive to the learned partitioning in a similar manner as the total energy. This property can be observed in Figure~\ref{fig:prop_dist}, where the distributions of forces acting on e.g. hydrogen and carbon atoms show a similar agreement between models as do the molecular energies, but not the atomic energies.

\subsection{Adaptive Sampling with Multiple Replicas}
\label{subsec:3.4}

A potential problem of the adaptive sampling scheme is its serial nature. Currently, only one point of data is collected after each sampling period. Since the NNPs need to be retrained every time the reference data set is extended, the resulting procedure can become time consuming in its later stages, especially for large and flexible molecules (e.g. the tripeptide in Reference~\cite{Gastegger2017CS}).

\begin{figure}
	\centering
	\includegraphics[width=0.7\textwidth]{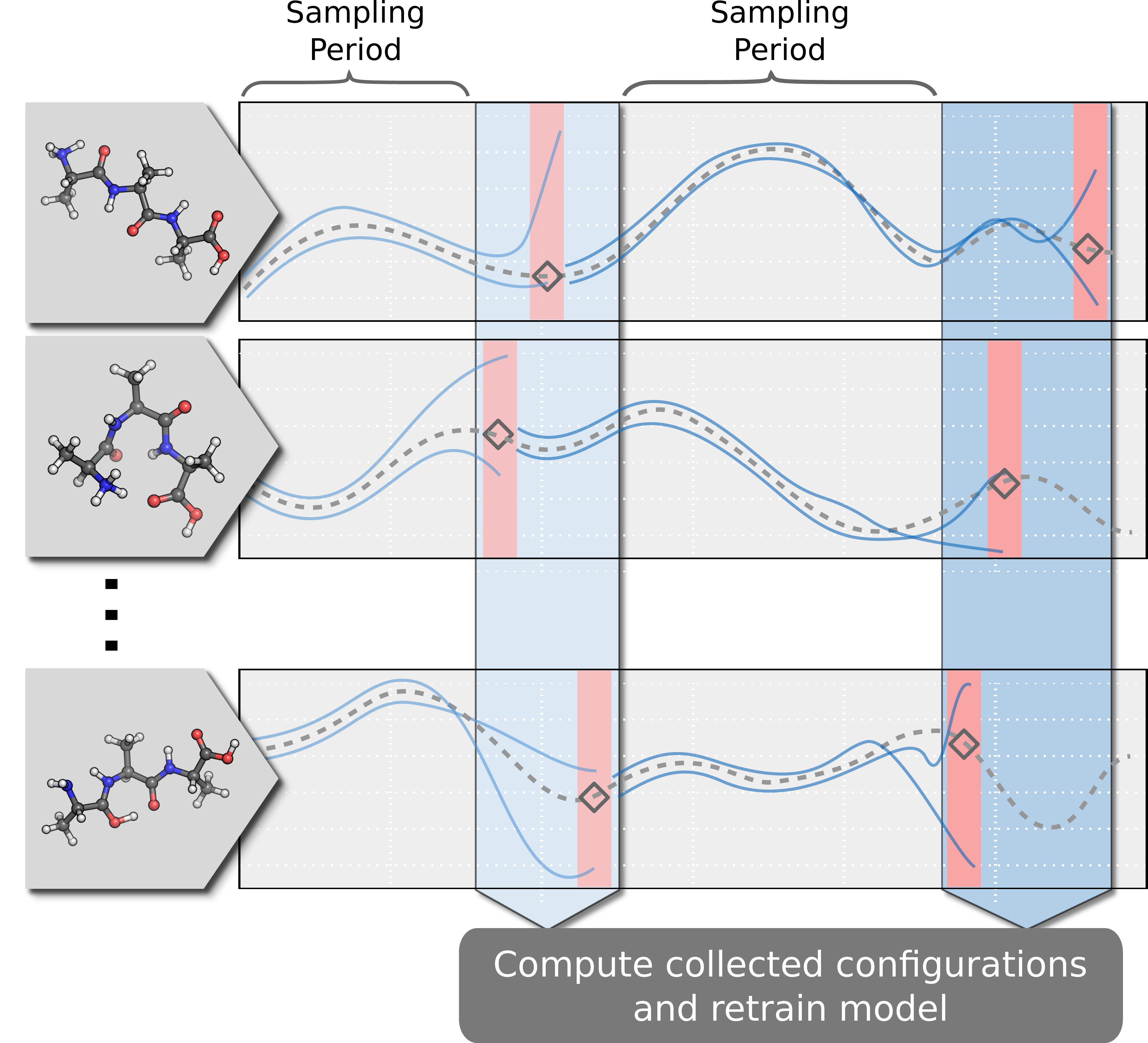}
	\caption{Parallel version of the adaptive sampling scheme. Individual adaptive sampling runs are carried out for different replicas of the system (e.g. different configurations). For each replica, configurations with high uncertainty are identified. Once samples have been collected for all replicas, reference computations are carried out and the NNP ensemble model is retrained. Afterwards, the replica simulations are continued with the new model. In this manner, the NNPs have to be retrained less frequently and different regions of the PES can be explored more efficiently.}
	\label{fig:msamp}
\end{figure}

This problem can be overcome by introducing a parallel version of the adaptive scheme, as outlined in Figure~\ref{fig:msamp}.
Instead of simulating only a single system at a time, multiple sampling runs are performed in parallel.
These independent simulations can be replicas of the system under various conditions (e.g. different temperatures), a range of conformations or even different molecules.
Sampling is once again carried out until divergence is reached for all parallel simulations. The high uncertainty configurations are then computed with the reference method and added to the training data.
This setup reduces the frequency with which NNPs need to be retrained, while at the same time improving PES exploration.
A potential drawback of this scheme is, that the collection of data points introduces periods of unproductivity, where some simulations are already finished while others are still running. However, this effect is negligible in praxis due to the high computational speed of the NNP models.


\section{NNPs for Molecular Dynamics Simulations}
\label{sec:4}

Due to their combination of high accuracy and computational efficiency, NNPs are an excellent tool to accelerate MD simulations.
A particularly interesting application is the computation of molecular spectra via the Fourier transform of different time autocorrelation functions \cite{Thomas2013PCCP}.
Depending on the physical property underlying the autocorrelation function, different types of spectra can be obtained.
One example are molecular infrared spectra, which can be modeled according to the relation:
\begin{equation}
I_\mathrm{IR} \propto \int^{+\infty}_{-\infty} \langle \dot{{\bm{\mu}}}(\tau) \dot{\bm{\mu}}(\tau+t) \rangle_\tau ~e^{-i\omega t} \mathrm{d}t,
\label{eq:dipauto}
\end{equation}
where $\dot{\bm{\mu}}$ is the time derivative of the molecular dipole moment, $\tau$ is a time delay, $\omega$ is the vibrational frequency and $t$ is the time.

The simulation of infrared spectra poses a particular challenge for machine learning techniques.
Due to the dependence of Equation~\ref{eq:dipauto} on $\dot{{\bm{\mu}}}$, a reliable model of the molecular dipole moment $\bm{\mu}$ is needed in addition to the PES description provided by conventional NNPs.
In the next sections, we will explore various aspects and the potential pitfalls associated with such models.

\subsection{Machine Learning for Molecular Dipole Moments}
\label{subsec:4.1}

A straightforward way to model dipole moments in the context of NNPs is to train individual atomic networks to reproduce quantum chemical partial charges. The molecular dipoles can then be obtained via the point charge model given in Expression~\ref{eq:atmu}, where the $q_i$ are now replaced by the environment-dependent network predictions $\tilde{q_i}$.
A similar strategy was e.g. used to model long-range electrostatic energies in Reference~\cite{Morawietz2012JCP}.

\begin{figure}
	\centering
	\includegraphics[width=0.9\textwidth]{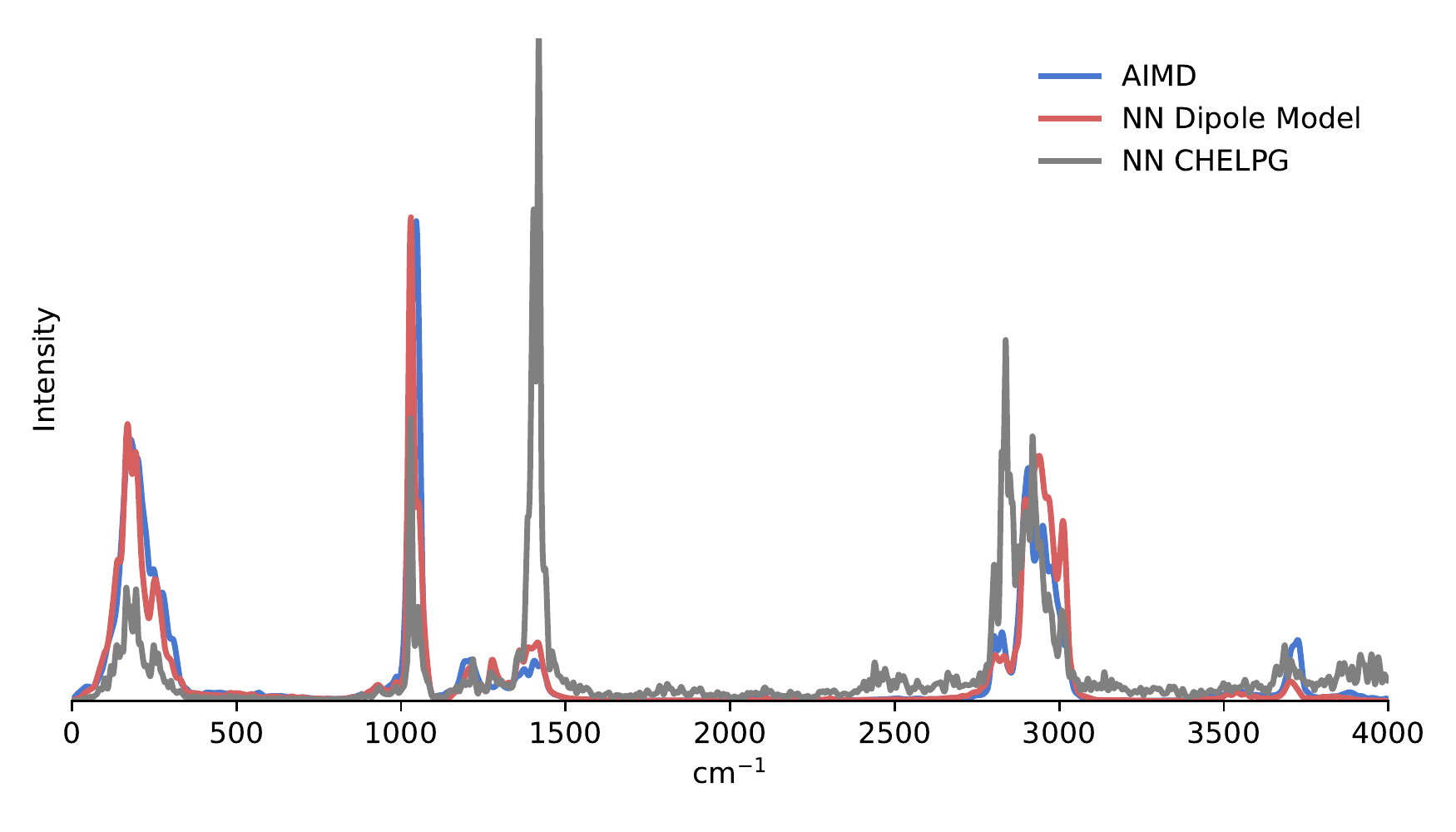}
	\caption{Infrared spectra of a methanol molecule in the gas phase computed via \emph{ab initio} molecular dynamics (blue), as well as machine learned molecular dynamics using the dipole moment model introduced above (red) and a neural network model trained on CHELPG partial charges (gray). While the dipole moment model shows good agreement with the reference, the CHELPG model leads to erratic trends in peak positions and intensities.}
	\label{fig:ir}
\end{figure}

However, such a model suffers from the inherent inhomogeneity of the various charge partitioning schemes available for electronic structure methods. The predicted partial charges can differ dramatically between schemes and some of them fail at reproducing molecular dipole moments entirely \cite{Cramer2004}.
Even when considering only those methods which yield partial charges consistent with the molecular dipole moment,  a strong dependence on the type of partitioning can still be observed.
Hirshfeld charges \cite{Hirshfeld1977TCA}, for example, appear to work well in the setup described above, as was demonstrated in Reference~\cite{Sifain2018JPCL}.
Charges fit to the electrostatic potential (e.g. CHELPG \cite{Breneman1990JCC}) on the other hand prove to be more problematic.
To illustrate the issue at hand, Figure~\ref{fig:ir} shows the MD IR spectrum of single methanol molecule computed with a partial charge model based on the CHELPG method in comparison to the electronic structure reference.
The partial charge spectrum shows several marked differences from the reference. 
Small artificial peaks are introduced at 2100 and 3900~\icm\ respectively. Moreover, the intensity of several peaks (e.g. at 1400~\icm\ and 2800~\icm) is reproduced incorrectly.
The most likely reason for these issues is the fitting procedure used to determine this particular type of reference charges.
Since an independent least squares optimization is carried out for every molecular configuration, the obtained partial charges are not necessarily continuous with respect to incremental changes in the local environment of each atom. This makes it harder for the atomistic networks to learn a consistent charge assignment, leading to the erroneous behavior observed above.

A better approach is to incorporate the point charge model into the atomistic NNP architecture in the form of a dipole aggregation layer, as described in subsection~\ref{subsec:2.2} and Reference \cite{Gastegger2017CS}. Instead of fitting to arbitrary partial charges, the model can now be trained directly on the molecular dipole moments, which are quantum mechanical observables. In this manner, the need for choosing an appropriate partitioning scheme is eliminated.
The inherent advantage of such a model can be seen in Figure~\ref{fig:ir}, where it accurately reproduces the quantum chemical reference, although trained on the same set of configurations as the partial charge model.

\subsection{Latent Partial Charges}
\label{subsec:4.2}

A special feature of the above model is that it offers access to atomic partial charges. These charges are inferred by the NNP model based on the molecular electrostatic moments in a purely data driven fashion.
Moreover, the charge models obtained with the above partitioning scheme depend on the local chemical environment of each atom. Hence, the charge distribution of the molecule can adapt to structural changes.
Considering that partial charges are one of the most intuitive concepts in chemistry, the NNP latent charges represent an interesting analysis tool, e.g. for rationalizing reaction outcomes.
In the following, we investigate how well the charges derived from the above dipole model agree with basic chemical insights.

\begin{figure}
	\centering
	\includegraphics[width=0.9\textwidth]{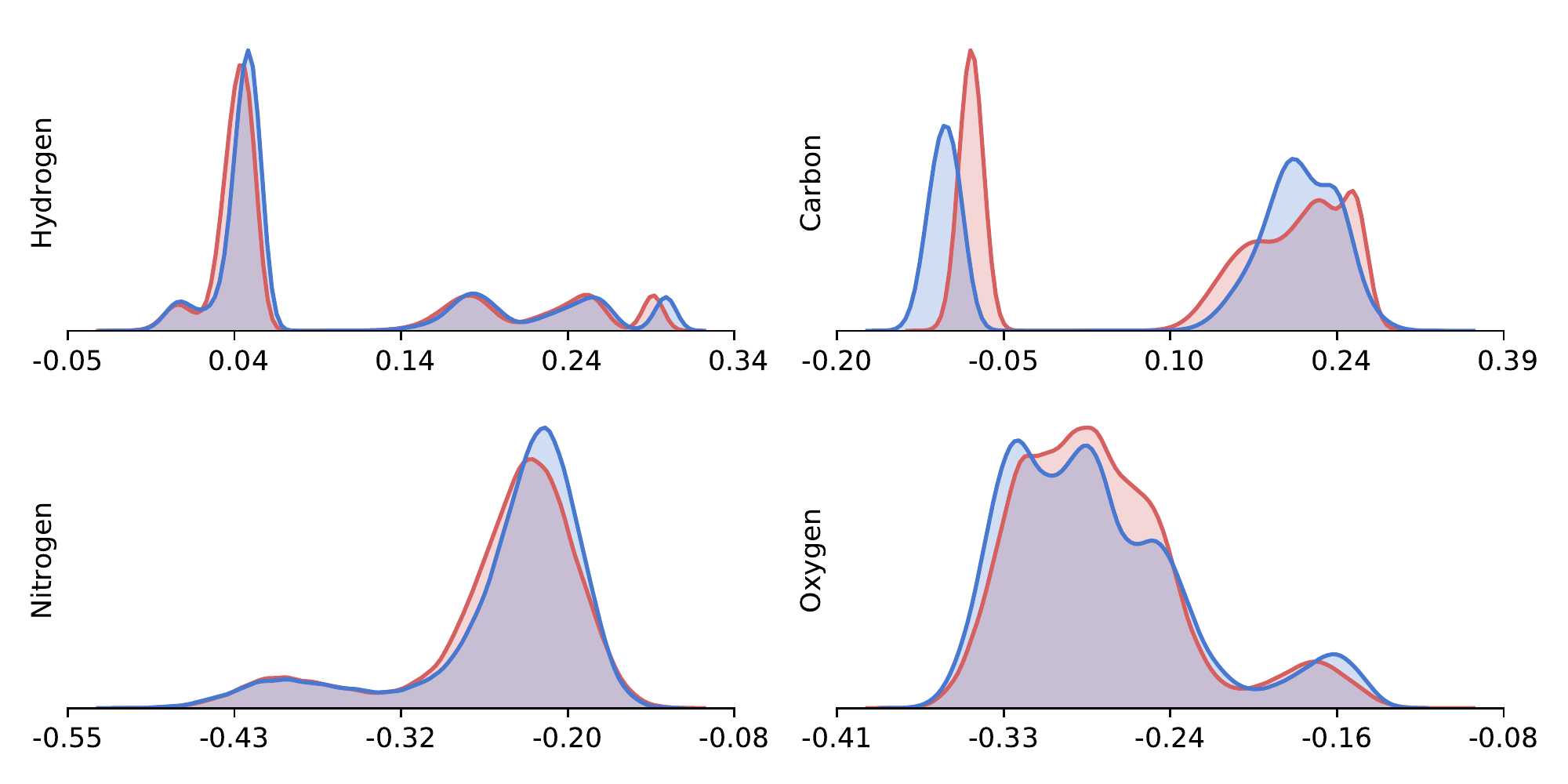}
	\caption{Distribution of atomic partial charges predicted for the chemical elements present in the alanine tripeptide obtained with two dipole models (blue and red). Although differences between the two models are still present, the atomic charge distributions are better conserved than the atomic energies.}
	\label{fig:charged}
\end{figure}

One potential problem of atomistic properties (e.g. energies and charges) obtained via specialized aggregation layers is the extent with which the partitioning varies between different models.
A good example are the atomic energies predicted by the tripeptide NNPs shown in Figure~\ref{fig:prop_dist}.
Although the total energies agree well, the partitioning into atomic contributions is highly non-unique.
Such a behavior is detrimental if the latent contributions should serve as an analysis tool.
In order to investigate whether this phenomenon is also observed for the latent partial charges, a similar analysis is performed for two dipole moment models of the alanine tripeptide. The resulting partial charge distributions are compared in Figure~\ref{fig:charged}.
The latent charges obtained with the dipole model are significantly better conserved than the atomic energies and only small deviations are found between different NNPs. The reason for this behavior is the geometry dependent term present in Equation~\ref{eq:atmu}, which introduces an additional constraint into the partitioning procedure.
These results are encouraging and demonstrate that the NNP partial charges are indeed capable to capture aspects of the chemistry underlying a system. However, care should be taken when using the latent charges as a direct replacement of their quantum chemical counterparts, as the resulting partitionings -- although well behaved -- are still not uniquely determined. This can lead to undesirable effects when they are e.g. used to model long-range electrostatic interactions without further processing, as it can introduce inconsistencies into the predicted model energies and forces \cite{Yao2018CS}.

\subsection{Electrostatic Potentials}
\label{subsec:4.3}

Having ascertained the general reliability of the charge model, we now study how well the latent charge assignments agree with the predictions of electronic structure methods.
In order to illustrate and compare different molecular charge distributions, we use partial charges to construct approximate electrostatic potentials (ESPs) of the form:
\begin{equation}
E(\mathbf{r}_0) = \sum^N_i \frac{{q}_i q_0}{||\mathbf{r}_i-\mathbf{r}_0||},
\end{equation}
where ${q}_i$ and $\mathbf{r}_i$ are the partial charge and position vector of atom $i$.
$\mathbf{r}_0$ is the position of a probe charge $q_0$, which was set to $q_0=+1$ in all experiments.

\begin{figure}
    \newcolumntype{S}{>{\centering\arraybackslash} m{0.02\textwidth} }
    \newcolumntype{T}{>{\centering\arraybackslash} m{0.45\textwidth} }
    \begin{tabular}{STT}
    \rotatebox{90}{Latent Charges} & \includegraphics[width=0.43\textwidth]{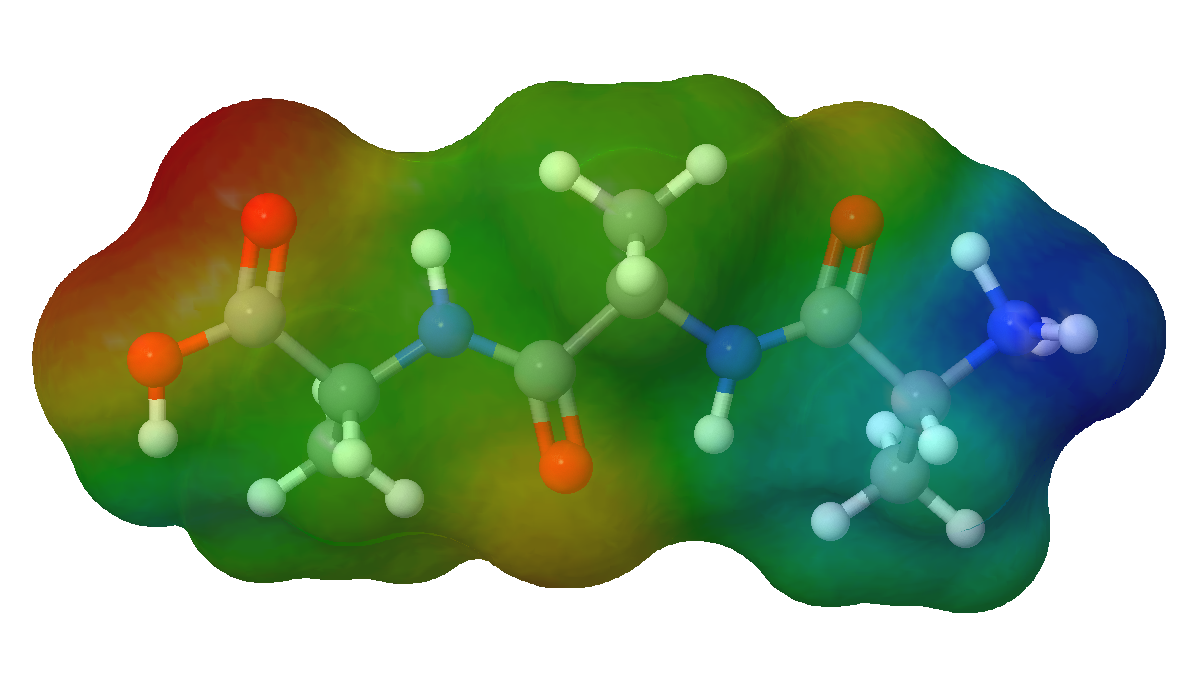} & \includegraphics[width=0.43\textwidth]{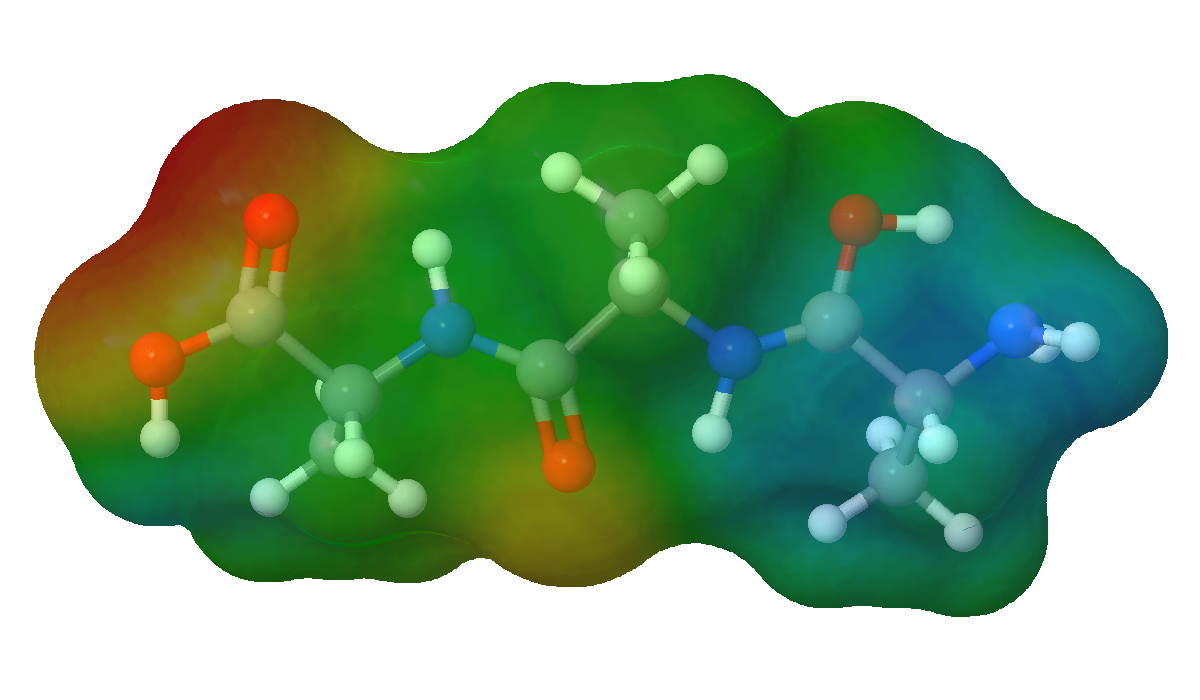} \\
    \rotatebox{90}{Hirshfeld} & \includegraphics[width=0.43\textwidth]{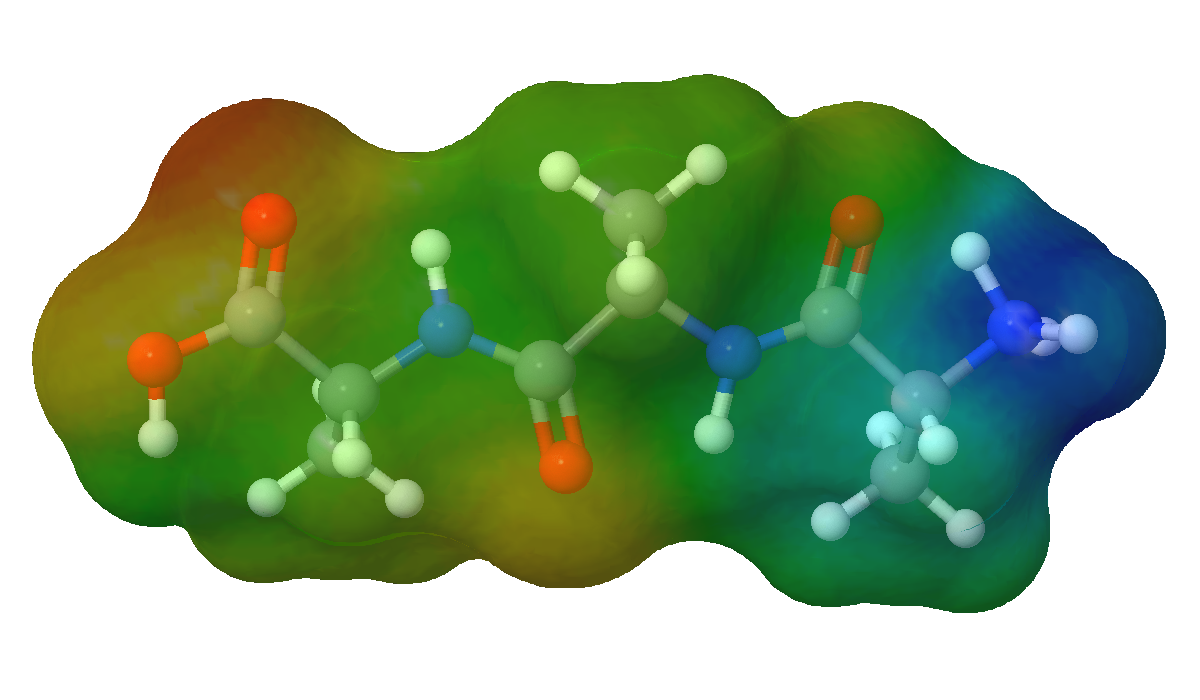} & \includegraphics[width=0.43\textwidth]{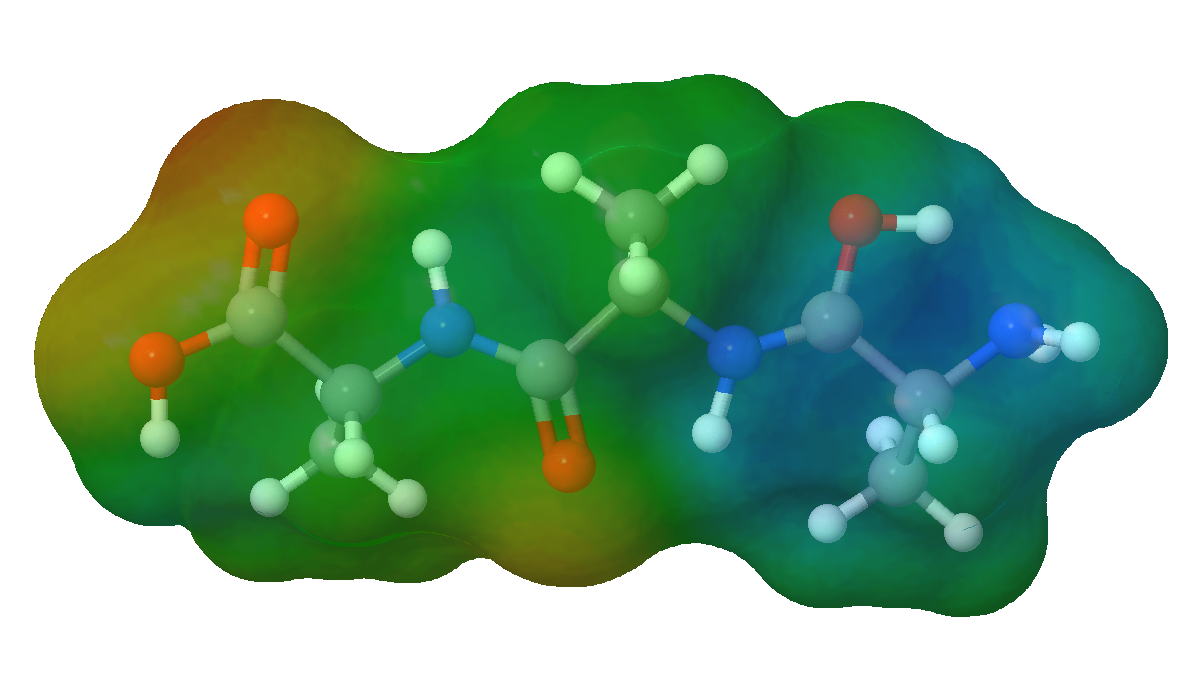} \\
    \end{tabular}
    \centering
    \caption{Electrostatic potential surfaces of the alanine tripeptide based on Hirshfeld partial charges and latent partial charges yielded by the dipole model. The left-hand side shows a configuration protonated at the N-terminal \ce{NH3+} group, whereas the proton is situated on the adjacent carbonyl group in the right-hand side structure. Regions of negative charge are depicted in red, positively charged regions in blue.}
    \label{fig:espnh3}
\end{figure}

Figure~\ref{fig:espnh3} shows the pseudo ESPs obtained with latent and Hirshfeld partial charges. 
The latter have been chosen for their general reliability and widespread use. To assess, how well the latent predictions of the dipole model capture the charge redistribution associated with changes in the molecular geometry, two configurations of the protonated alanine tripeptide are modeled, with a hydrogen transferred from the N-terminal \ce{NH3+} group to the neighboring carbonyl moiety.

In all cases, good agreement is found between the charge distributions predicted by the dipole moment model and the electronic structure equivalent. The latent charges are able to account for several important features, such as the localization of the positive charge of the molecule at the N-terminal \ce{NH3+} moiety in the first configuration, as well as its relocation towards the interior of the molecule after proton transfer.
The model also accounts for the regions of negative charge expected for the carbonyl and carboxylic acid groups. These findings are remarkable insofar, as all this information on the electronic structure of the molecule is inferred purely from the global dipole moments, demonstrating the power of the partitioning scheme.

\subsection{Geometry Dependence of Latent Charges}
\label{subsec:4.4}

\begin{figure}
	\centering
	\includegraphics[width=0.7\textwidth]{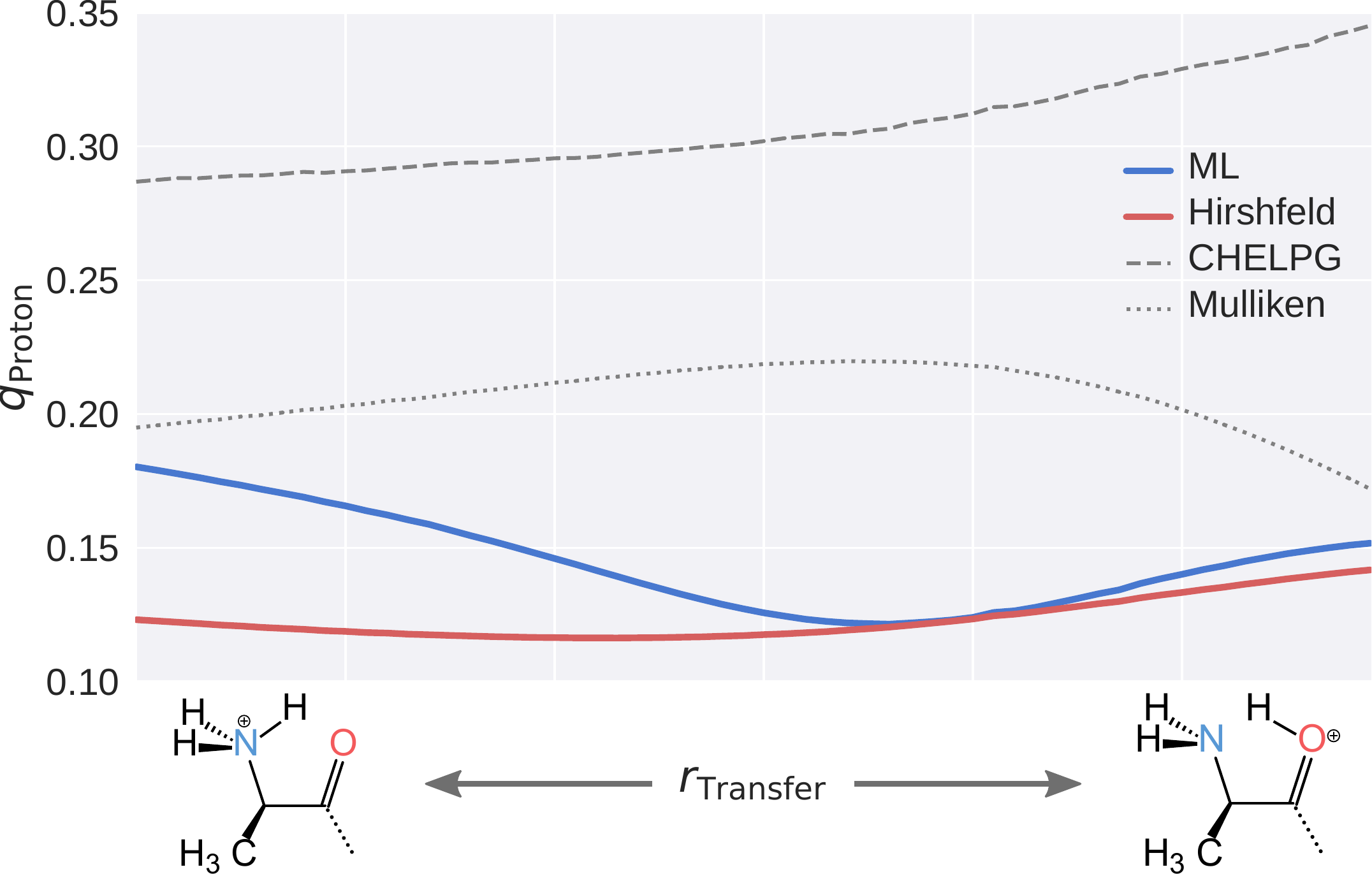}
	\caption{Changes in the partial charge of the proton during different stages of the proton transfer event. Shown are charges computed via different conventional charge partitioning schemes (Hirshfeld, CHELPG and Mulliken), as well as the latent charges predicted by the ML model.}
	\label{fig:qproton}
\end{figure}

A final analysis is dedicated to the behavior of the latent dipole model charges under changes in the local chemical environment.
As an example, we study the evolution of the partial charge of the proton during the proton transfer event occurring in the alanine tripeptide.
Figure~\ref{fig:qproton} shows the NNP partial charge attributed to the proton plotted against the reaction coordinate. The curves for Hirshfeld, Mulliken \cite{Mulliken1955JCP} and CHELPG charges are included for comparison.
Several interesting effects can be observed.

First, the dipole model curve exhibits a minimum close to the transition state of the transfer reaction.
Since the latent charges can be seen as a proxy of the local electron density, this result can be interpreted as follows:
At the initial and final stages of the transfer, the positive charge is located mainly at the proton itself. However, during the transfer and especially close to the transition regions, electron density is shared between the three participating atoms (O, N and proton). Hence, the positive charge is reduced for these configurations.
This finding serves as an additional demonstration for the efficacy of the latent charge model. Although originally only conceived to model dipole moments, it is able to provide insights directly related to the electronic structure of the molecule at an atomistic resolution.

Second, Figure~\ref{fig:qproton} illustrates the inherently different behavior found for various charge partitioning schemes. The Hirshfeld charges show a qualitatively similar curve to the machine learned charge model and are well behaved in general, supporting the results reported in Reference~\cite{Sifain2018JPCL}.
Mulliken and CHELPG charges on the other hand show completely different trends. The former are generally known for their unreliability, hence the result is little surprising \cite{Cramer2004}.
The counterintuitive behavior of the CHELPG charges serves as an additional confirmation for the effects observed in the methanol spectrum shown above (Figure~\ref{fig:ir}).
Given this general discrepancy between various partitioning schemes, the charges derived via the dipole moment model constitute a viable alternative:
They reproduce molecular dipole moments accurately, are derived directly from quantum mechanical observables and capture the influence of structural changes on the molecular charge distribution.

\section{Conclusion}
\label{sec:5}

We have presented how molecular dynamics (MD) simulations can benefit from machine learning (ML) potentials and provided some background for the implementation of this ML-MD approach. The first challenge during such a task is to efficiently gather enough training data in order to create a converged potential. An adaptive sampling scheme can serve for this purpose and the efficiency can be improved when using a) an ensemble of neural networks, b) an adequate uncertainty measure as selection criterion, and c) multiple replicas to parallelize the sampling.

As an ultimate test, experimental observables need to be calculated and compared to actual experimental results. In our case, infrared spectra are simulated, for which the neural networks do not only need to learn potentials and forces but also dipole moments and their atomistic counterparts, the atomic partial charges.
If the latter are plotted in a geometry-dependent manner, e.g., along a reaction coordinate, these machine-learned charges provide insights directly related to
the electronic structure of the molecule at an atomistic resolution.
In this sense, machine learning can not only deliver potentials with supreme accuracy at compelling speed but also offer valuable insights beyond a simple prediction of quantities.

Despite this positive picture, many challenges remain, e.g., generalizing the machine learning models -- ideally for all possible substances, different kinds of molecules and materials alike -- or extending the range of properties to be learned. Possibly the biggest challenge is however to find a universally valid electronic structure method necessary for the generation of high-fidelity training data.

\begin{acknowledgement}
M.G. was provided financial support by the European Unions Horizon 2020 research and innovation program under the Marie Sk\l{}odowska-Curie grant agreement NO 792572. The computational results presented have been achieved in part using the Vienna Scientific Cluster (VSC). We thank J. Behler for providing the RuNNer code.
\end{acknowledgement}

%
%
%
%


\end{document}